# Effect of La Doping on the Crystal Structure, Electric, Magnetic and Morphologic Properties of the BSCCO System

*Vivian Delmute Rodrigues[a], Gisele Aparecida de Souza[a], Claudio Luiz Carvalho[a], Rafael Zadorosny[a]**

[a]*Faculdade de Engenharia, Universidade Estadual Paulista - UNESP, 15385-000, Ilha Solteira, SP, Brasil.*

Studies of the doping process can provide a better understanding of the superconducting mechanisms in cuprous oxide materials. In this work, we studied the doping effects on the crystal structure, electric, morphologic and magnetic properties of the BSCCO system with the nominal composition $Bi_{1.6}Pb_{0.4}Sr_{(2-x)}RE_xCa_2Cu_3O_{(10-\delta)}$. Here, the rare earth element (RE) was replaced by La in the sites of Sr. The x was ranged from 0.0 to 2.0, in steps of 0.5. The samples were prepared based on Pechini's method. The resulting powder was pressed at room temperature and the pellets were submitted to several heat treatments. The characterizations confirm the La in the sites of Sr however; the superconducting properties of the sample were not improved.

**Keywords:** *Doping, La, BSCCO, superconductivity.*

## 1. Introduction

Since the discovery of the superconducting copper-oxides, the doping or the substitution processes have been used to improve some of their properties as the critical current density[1-3] and some applications[4,5].

BSCCO (Bi-Sr-Ca-Cu-O) is one of the most studied superconducting systems. The $T_C s$ of such system ranged from 7 K to 110 K and are directly related to the four possible superconducting phases, i.e., $Bi_2Sr_2CuO_6$ (Bi-2201), $Bi_2Sr_2CaCu_2O_8$ (Bi-2212), $Bi_2Sr_2Ca_2Cu_3O_{10}$ (Bi-2223) and $Bi_2Sr_2Ca_3Cu_4O_{12}$ (Bi-2234)[2,6-8]. Some goals of the BSCCO system are: (i) more acceptable, ecologically speaking, when compared with systems that have Tl and Hg in their compositions; (ii) presents a high critical current density and (iii) the $H_{C2}$ is higher than of the other systems. All these features make such material promising for future applications[6,7].

There are several studies that analyze various doping elements in different sites of BSCCO, such as rare earth elements[4,9-18,23,27,29], oxides as the $Cr_2O_3$[26], and some alkali metals and transition elements (Na, Li, Ba, Zn, Y, V, Fe, Hg, Pb)[1,6,9,19-22,24,25,28,29,30]. Those studies have shown changes in crystal structure, electric, morphologic and magnetic properties above a certain dopant content[4,5,9-24,31].

However, the specific methodologies used in the doping processes remain unclear. The doping with Pb in the Bi sites is often performed to obtain thermodynamic stability of the Bi-2223 phase, and some works suggest that the presence of Pb enhances the connections between grains[1,9,25].

In this work, we studied the doping effects on the crystal structure, electric, magnetic and morphologic properties of the BSCCO system. The samples were prepared with nominal composition $Bi_{1.6}Pb_{0.4}Sr_{2-x}La_xCa_2Cu_3O_{10+\delta}$, where the rare earth element lanthanum (La) was inserted into the Sr sites. The x parameter was ranged from 0 to 2.0 in steps of 0.5.

## 2. Material and Experimental Procedure

The precursor solutions were prepared based on the Pechini´s method[29,32,33]. The used reagents were: $Bi_2CO_5$, $SrCO_3$, $CaCO_3$, $CuCO_3.Cu(OH)_2$, $2PbCO_3.Pb(OH)_2$ (Vetec), $La_2(CO_3)_3.H_2O$ (Apache Chemicals), $C_6H_8O_7$, $HNO_3$, $C_2H_8N_2$, and $C_2H_6O_2$ (Vetec and Synth).

The metallic carbonates (MC) were dissolved in an aqueous solution of citric acid ($C_6H_8O_7$) (CA) in a ratio of 3:1 (MC:CA). After that, a polyol, in this case, the ethylene glycol ($C_2H_6O_2$) was added to obtain a better distribution of the cations in the polymeric structure. We used the proportion of 3:2 mol of ethylene glycol:citric acid and then, the solution was heated at 80ºC-90ºC to be dried. Consequently, heat treatments at 200ºC/10h, 400ºC/6h, 600ºC/6h, 800ºC/6h and 810ºC/68h, with a heating rate of 2ºC/min were done. The resulting powder was grounded at each heat treatment interval and then an amount of (1.2000±0.0001) g was separated and pressed uniaxially. The thickness of all pellets was (1.03±0.01) mm. The pellets were calcined in an air atmosphere at 810ºC/31h with a heating rate of 2ºC/min.

* e-mail: rafazad@yahoo.com.br

The samples were characterized by X-ray diffraction (XRD) with Cuα radiation (λ= 1,542 Å). The measurements ranged from 2θ = 5° to 70° in steps of 0.2°. The dc four-probe method was used to the electric characterization (77K≤T≤300K). For the morphology characterization, it was used a scanning electron microscopy (SEM) coupled with an energy dispersive spectroscopy X-ray (EDX) detector. Ac magnetic susceptibility and Raman Spectroscopy (with excitation at λ = 514.5 nm) were also used. The samples were labeled as La(0), La(0.5), La(1.0), La(1.5) and La(2.0), according to the La content in the nominal composition.

## 3. Results and Discussion

Fig. 1 shows the XRD patterns for all samples. In the analysis it were used the follow references: JCPDS 80-2029 for Bi-2212, JCPDS 80-2030 and JCPDS 42-0450 for Bi-2223, JCPDS 39-0283 for Bi-2201, JCPDS 50-0686 for $(Sr_xCa_{2-x})CuO_3$ and JCPDS 76-2128 for $Ca_2PbO_4$. We observed that most of the La(0) peaks were due to Bi-2212 and Bi-2223 phases with some of them overlapped. We also observed the presence of other phases and sub products such as Bi-2201 and $Ca_2PbO_4$. All peaks related to Bi-2212 and Bi-2223 were indexed based on an orthorhombic symmetry.

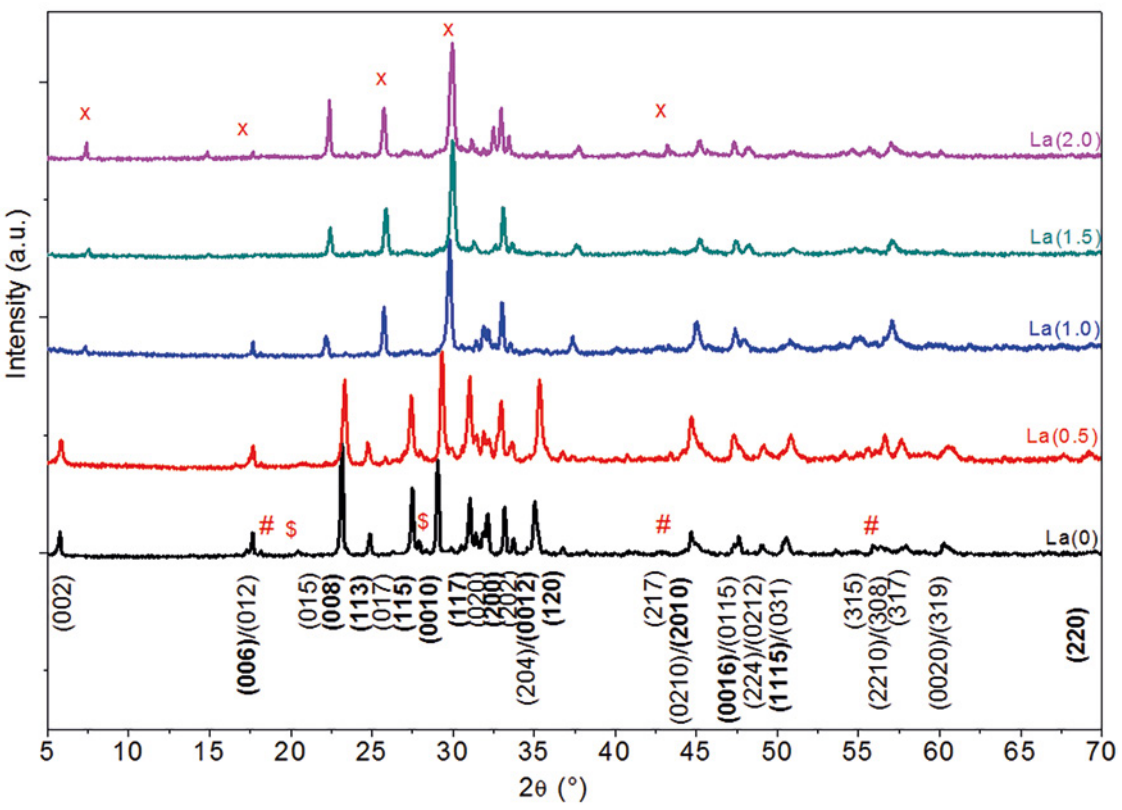

**Figure 1.** XRD of samples with different doping contents. The figure shows that the indexed peaks which are related to Bi-2212 and Bi-2223 superconducting phases. The subproducts were (#) $Ca_2PbO_4$, ($)(SrxCa2-x)$CuO_3$ and (x)Bi-2201.

When compared with La(0), the La(0.5) did not show significant changes in its XRD with the presence of 2223 and 2212 phases. The other doped samples showed changes in the positions of the peaks when compared to the La(0) and La(0.5). For contents greater than x ≥ 1.0, the samples were indexed as containing mostly the Bi-2201 and Bi-2212 phases. The peaks near 7.3° and 25.7° are associated with the Bi-2201 phase. This suggests that as the dopant content is increased, it allows the formation of the lower $T_C$ phases such as Bi-2201 and Bi-2212, and not improve the stability of the Bi-2223 phase. According to Abbas *et al.* [34], the formation and stability of the Bi-2223 phase can be modified by the addition or substitution of elements with different ionic radius and bonding characteristics.

For a qualitative analysis, the relative volume fractions of Bi-[22(n-1)n] were estimated by Eq.(1), where [n= 1,2,3] indicates the phases 2201, 2212 and 2223. The peaks used in these calculations were chosen in the range of 2θ= 5-50° as done in the work of Taghipour *et al.*[35].

$$\mathrm{Bi}[22(n-1)n](\%) \approx \frac{\sum I_{\mathrm{Bi}\text{-}[22(n-1)n]}}{\sum I_{\mathrm{Bi}\text{-}2223} + \sum I_{\mathrm{Bi}\text{-}2212} + \sum I_{\mathrm{Bi}\text{-}2201}} \times 100 \quad (1)$$

Here, I is the intensity of the peaks of each phase. Table 1 shows the relative volume fractions. In general, the substitution of La in Sr sites increase the formation of the BSCCO phases, which are associated with the lower values of $T_C$. We can also infer that the doping process with La avoids the formation of the Bi-2223 phase.

**Table 1.** Relative volume fractions of the Bi-2223, Bi-2212 and Bi-2201 phases for all fabricated samples.

| Phase | La(0) | La(0.5) | La(1.0) | La(1.5) | La(2.0) |
|---|---|---|---|---|---|
| Bi-2223 (%) | 87.0 | 64.6 | 41.6 | 27.5 | 34.1 |
| Bi-2212 (%) | 10.0 | 33.0 | 44.6 | 33.9 | 35.1 |
| Bi-2201 (%) | 3.0 | 2.4 | 13.8 | 38.6 | 30.8 |

To obtain the lattice parameters, Eq. (2), we assumed an orthorhombic symmetry corresponding to the peaks associated with Bi-2212 in the XRDs[36]:

$$\frac{1}{d^2} = \frac{h^2}{a^2} + \frac{k^2}{b^2} + \frac{l^2}{c^2}. \quad (2)$$

Here *h,k,l* are the Miller indices; *a*, *b* and *c* are the lattice parameters and *d* is the distance between adjacent planes in the set (*hkl*)[36]. The (*00l*) directions were used to calculate *c*, the (200) direction were used to obtain *a*, and (*hkl*) to calculate *b*. Fig. 2 shows the lattice parameters for all samples.

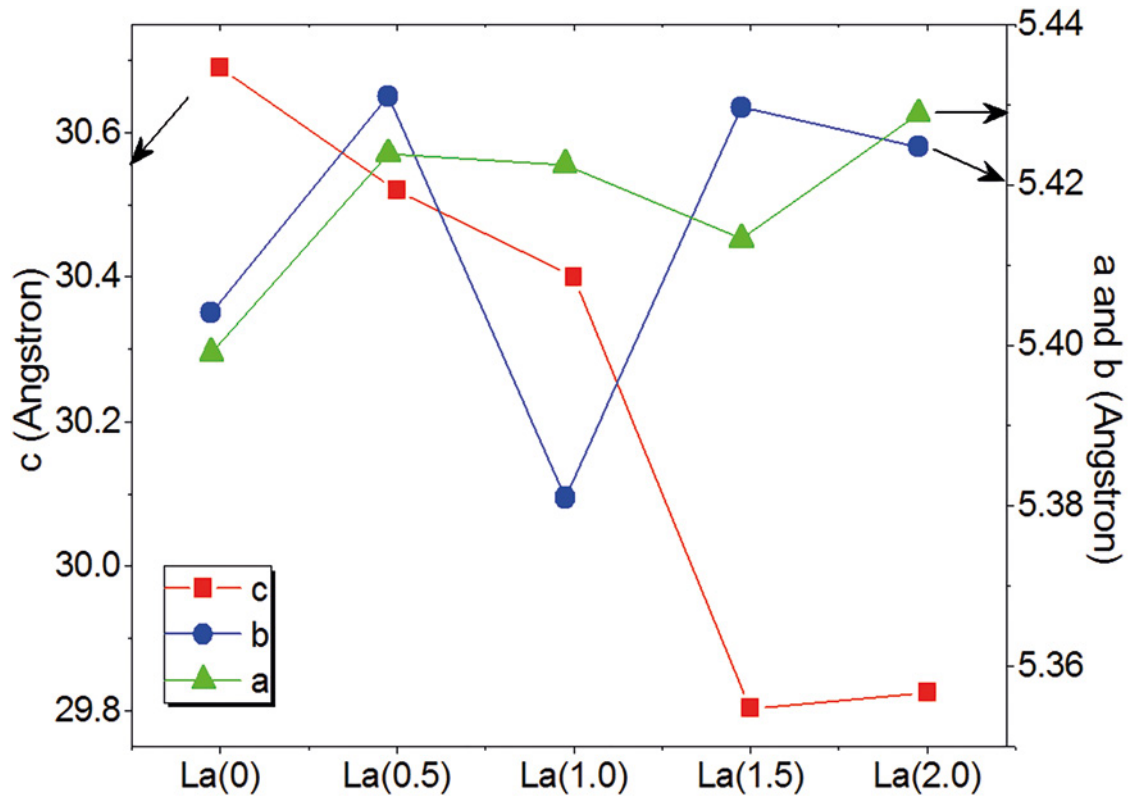

**Figure 2.** Evolution of the lattice parameters with increasing dopant content.

The lattice parameter slightly increases in the doped samples (comparing with La(0)). The parameter *b* also increases, except for the La(1.0). The parameter *c* decreases for all doped samples, which suggests that La occupies the Sr site. Such conclusion is reasonable since the ionic radius of La (1.06 Å)[10] is lower than that of the Sr (1.12 Å)[10].

The influence of La in the formation of crystallites is analyzed by calculating the typical sizes of these specimens by the Scherrer equation. Those analyses focused on the peaks around the positions 17º, 29º, 33º and 45º which are associated with the plans (006), (0010), (200) and (0210)/(2010), for the Bi-2212 and Bi-2223 phases, respectively. From Table 2 is evident that the average size of the crystallite of sample La(0) is greater than that one of the other samples. As the average size of the crystallites is related to the crystallinity of the sample, i.e., larger crystallites are associated with more crystalline materials[36], the data suggests that the doping process affects the morphology of the material.

**Table 2.** Size of crystallites estimated from the Scherrer's equation. It were used the peaks at 2θ =17º, 29º, 33º and 45º for which the average sizes were evaluated.

| 2θ | La(0) (nm) | La(0.5) (nm) | La(1.0) (nm) | La(1.5) (nm) | La(2.0) (nm) |
|---|---|---|---|---|---|
| 17º | 60.4 | 68.1 | 68.1 | - | 64.3 |
| 29º | 50.3 | 34.8 | 38.1 | 32.4 | 24.3 |
| 33º | 102.3 | 46.8 | 52.8 | 53.1 | 32.9 |
| 45º | 54.0 | 43.6 | 36.4 | 22.8 | 10.5 |
| average size | 66.8 | 48.3 | 48.9 | 36.1 | 33.0 |

Fig. 3 shows the electric measurements of all samples. Both the La(0) and La(0.5) exhibit a superconducting transition, see in Fig. 3(a) to (c). The other samples exhibited a semiconductor-like behavior as shown in Fig. 3(d) to (f). La(0) presents two $T_C$s at 94.8 K and 100.6 K, which correspond to Bi-2212 and Bi-2223 phases, respectively. The $T_C$s of La(0.5) are 96.1 K and 101.3 K, which also correspond to Bi-2212 and Bi-2223 phases, respectively. The values of $T_C$s were considered as the temperature at the maximum local of the *dR/dT* curves (Fig.3(b) and (c)). In Table 3 is shown the $T_C$s values of all samples, where we can notice an improvement in the critical temperature of the sample La (0.5).

The transition width (ΔT) of La(0.5) is wider than that presented by La(0). In Fig. 3(b) and (c) are shown ΔT for La(0) and La(0.5) respectively. Those values were obtained in the half height of the peaks[7] been 6.4 K for La(0.5) and 2.7 K for La(0). The ΔT qualitatively indicates the purity degree of the material. The presence of impurities can significantly interfere with the electrical properties of the sample confining the percolation of the transport currents in the surface of the samples. We can conclude that the doped sample presents more impurities than the undoped one, which is in agreement with the XRD analysis.

The semiconducting-like behavior of the other samples was observed in the range of temperatures that were studied. This was probably due to the insertion of a large amount of dopant in the crystalline structure of the material. Then, changes in the charge reservoir layers with the substitution of $La^{3+}$ for $Sr^{2+}$ is induced and consequently, the amount of charge carriers in the superconducting planes is changed[4]. This behavior is usually due to the insulator-metal transition which occurs with other rare earth dopant elements, e.g., Y[5], Gd[12], Ho[14] and Dy[16] in the BSCCO system.

Fig. 4(a) and Fig. 4(b) show the IV curves of La(0) and La(0.5) samples, respectively. The measurements were carried out at 77K and the $I_C$ was determined by the 1μV/cm criterion. We considered the samples with a thickness of (1.03±0.01) mm and width of (4.00±0.01) mm. The distance between the voltage terminals was (2.00±0.01) mm. The $I_C$ is (5.5±0.1) mA and (1.1±0.1) mA, for La(0) and La(0.5) respectively. The related critical current density ($J_C$) is (0.133±0.004) A/cm$^2$ for La(0) and (0.030±0.003) A/cm$^2$ for La(0.5). These values show that $J_C$ of La(0.5) is about four times lower than that for La(0). In Ref. 4 was shown that both $J_C$ and $T_C$ can also be changed by reducing the charge carriers in the superconducting plans, i.e., substituting $Sr^{2+}$ by $La^{3+}$ the number of holes in the $CuO_2$ planes can change[4]. Thus, it is observed that the increase of the dopant content degrades the superconducting properties of the samples.

Fig. 5 shows the scanning electron microscopy (SEM) micrographs of the samples. Fig. 5(a) shows large grains with maximum dimensions of (12.8 x 13.2) μm$^2$. The plate-like morphology is characteristic of BSCCO systems. We also observed the presence of a bright area with a small amount of grains formed due to a possible partial melting during the sintering process. Fig. 5(b) shows a SEM micrograph of La(0.5), which presents a more homogeneous distribution of the grains when compared with the La(0) and with smaller sizes of about (5.2x3.6) μm$^2$.

Fig. 5(c) and (d) show the micrographs of La(1.0) and La(1.5), respectively. The size of the grains is smaller than those ones of the other samples, with dimensions of about (1.8 x 0.8) μm$^2$ and (1.2x1.6) μm$^2$, respectively. The micrograph of La(2.0) in Fig. 5(e) shows the presence of larger plate-like grains than that exhibited by La(1.0) and La(1.5), however, with a smaller size than that of La(0) and La(0.5) [(2.4x1.6) μm$^2$]. It was also observed that the grains sizes distribution of La(0), La(0.5) and La(2.0) are also more homogeneous than the other samples.

Table 4 shows the EDX of the samples. It is observed that some elements present slight changes when compared with the nominal composition $Bi_{1.6}Pb_{0.4}Sr_{2-x}RE_xCa_2Cu_3O_y$. The nominal composition of Cu for La(0) and La(0.5) is about 10% and 15% greater than expected, respectively, as well as the composition of La for samples La(0.5), La(1.0), La(1.5) and La(0.5), is about 14%, 11%, 15% and 11% greater than the nominal value, respectively. The Sr also presents deviations for La(1.0) and La(1.5) with values of about 13% and 15% greater than the expected ones. The nominal composition of the Pb is 22%, 18% and 32% greater than expected for La(1.0), La(1.5) and La(2.0) respectively. These changes may be associated with the loss of Pb due to long sintering times. Such consideration is also applied to the Bi[35]. Note that $La^{3+}$ seems to enter in the sites of $Sr^{2+}$ as the nominal composition of La increases while the Sr content decreases.

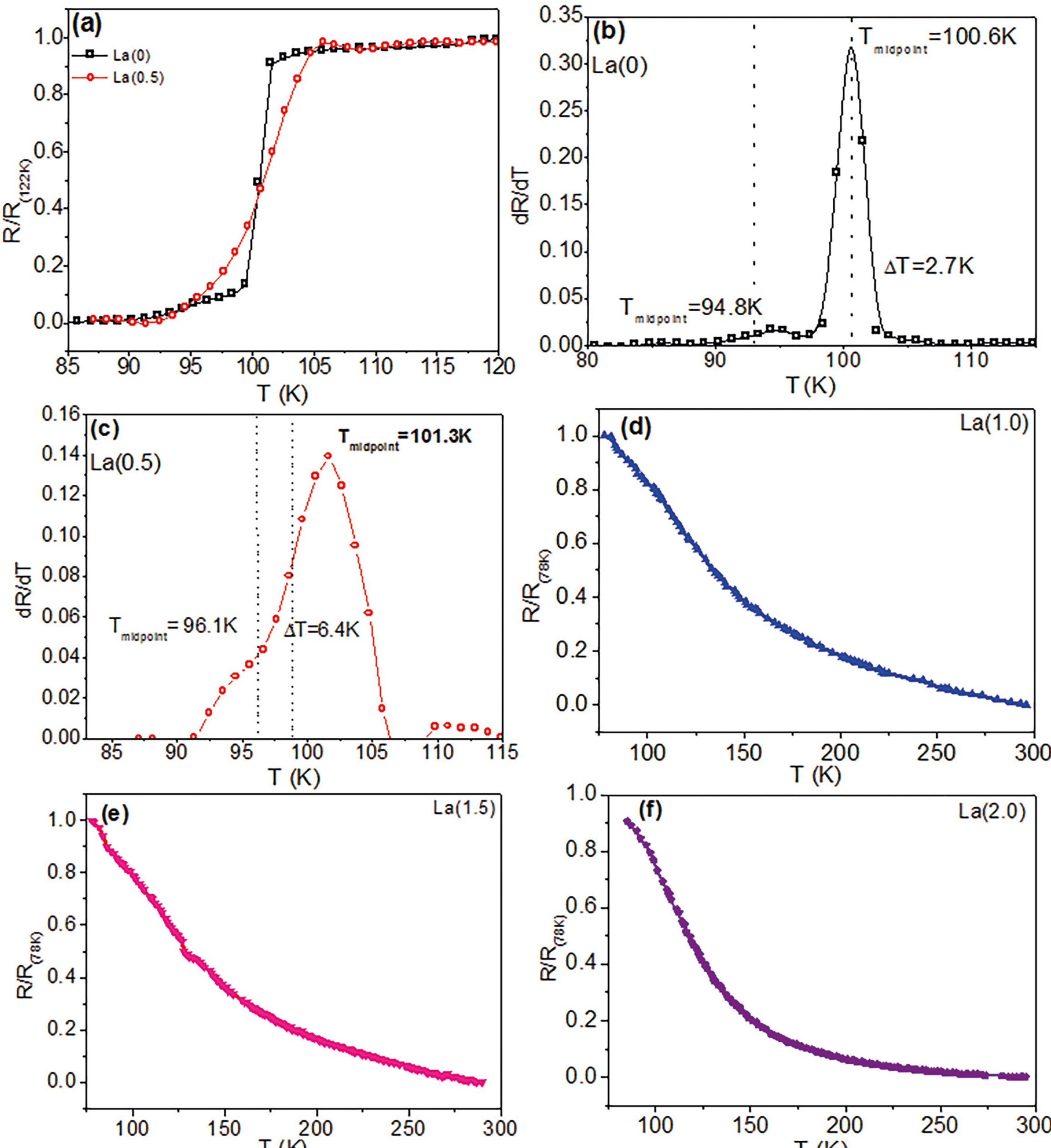

**Figure 3.** Electric measurements, RxT curves, using the dc four probe method of the samples: (a) La(0) and La(0.5), which present a superconducting transition; (b) and (c) are the (dR/dT)xT curves for the samples La(0) and La(0.5), respectively. The RxT curves of (d) La(1.0), (e) La(1.5) and (f) La(2.0), which exhibit a semiconducting-like behavior.

**Table 3.** Critical temperatures of the samples La(0) and La(0.5).

| Sample | Bi-2212 | Bi-2223 |
|---|---|---|
| La(0) | 94.8 K | 100.6 K |
| La(0.5) | 96.1 K | 101.3 K |

Measurements of ac magnetic susceptibility are shown in Fig. 6(a) and (b) for La(0) and La(0.5), respectively. For La(0) sample, in Fig. 6(a), two excitation fields were used, i.e., 0.01

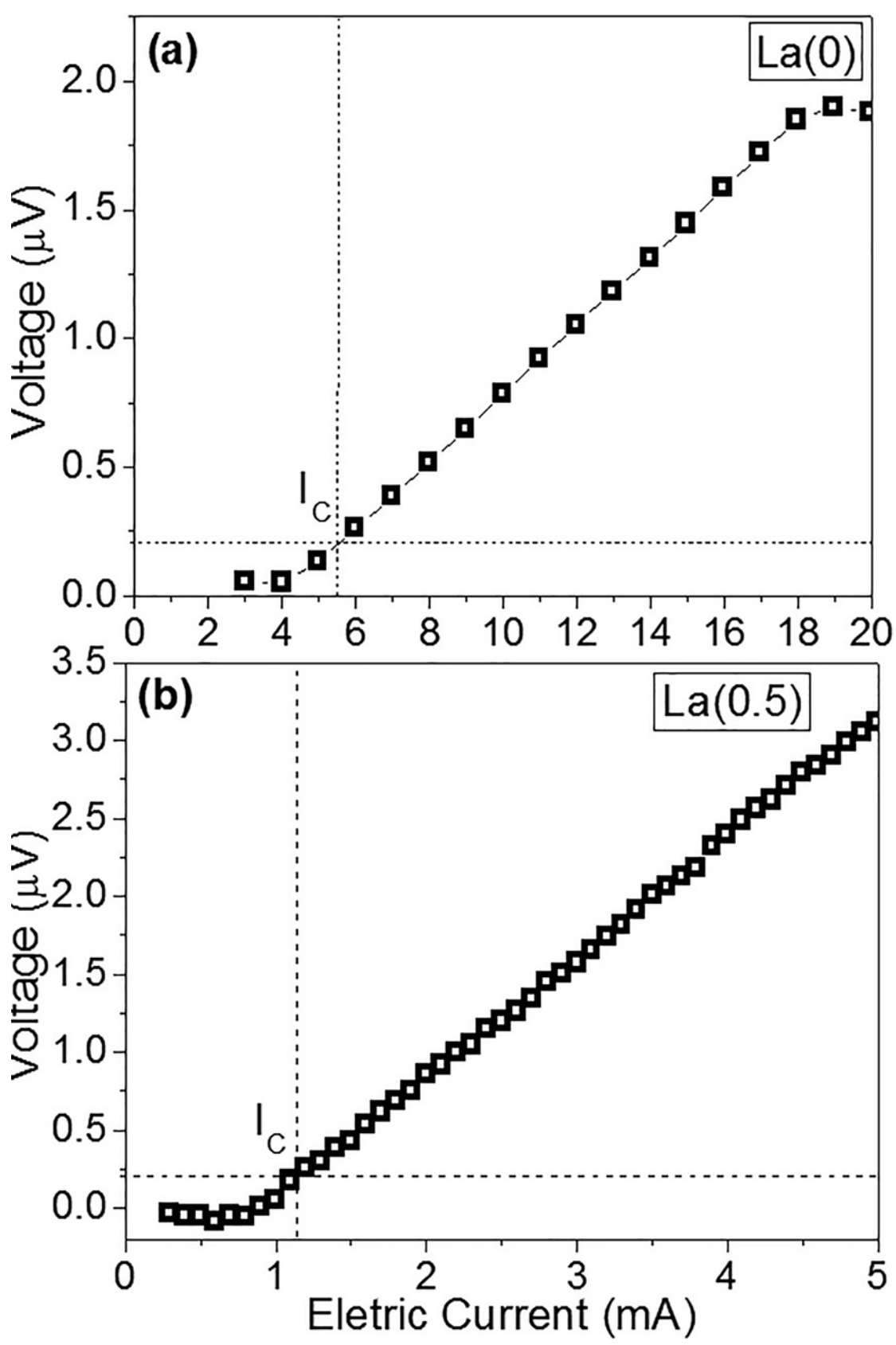

**Figure 4.** Voltage as a function of the electric current (IV curve) of (a) La(0) and (b)La(0.5). The dotted line shows the critical current (Ic) value.

Oe and 0.05 Oe and a frequency of 1 kHz. Two transitions are observed in that figure, the first one close to 75 K and the second one around 57 K. The lower temperature indicates that the weak-links are completely broken, i.e., there are no intergranular currents. The transition at 75 K is associated with the intrinsic diamagnetic shielding of the grains due to intragranular currents[37]. We can note that for 0.05 Oe, the second transition shifts to near 40 K and the first transition is not affected by the excitation field, which is a strong evidence that the second transition is related to the weak-links.

Fig. 6(b) shows the measurements of the La(0.5) sample, which exhibits a very degraded response in comparison with the response of La(0). We observed transitions at 7.1 K and 52.3 K, in the curve with the lowest excitation field (0.01 Oe). For the larger field (0.1Oe), the signal is smoother and the transition temperatures of the inter- and intragranular parts were observed at approximately 4.1 K and 57.0 K, respectively[38]. Those curves suggest that the superconducting properties of the La(0) sample are better than of La(0.5).

Comparing the ac susceptibility and the electric (*RxT*) measurements, we observed a large discrepancy between the values of $T_C$. Typically, the values of $T_C$ as well as $J_C$ are different for the electrical and magnetic measurements. As the electric measurements are associated with the percolation of the transport currents by the surface grains, the magnetic measurements provide an average value of the whole sample[38]. The mismatch between electric and magnetic measurements may be associated with the fabrication process of the samples, such as pressing the pellets, which can result in better contact between grains in the surface of the sample. The surface is also more sensitive to changes in the oxygen content, which causes a local degradation of the superconducting properties[38].

Fig. 7 shows the Raman spectra of all samples in the range from 150 to 1300 $cm^{-1}$. Raman spectroscopy is currently used in the investigation of the phonons and other low-energy excitations in high $T_C$ superconductors[38]. There are many works reporting on the assignment of Raman vibrational modes, but these studies are still controversial[39,40,41]. Table 5 shows the Raman spectra modes observed. These Raman frequencies are very similar to the frequency shown in the studies of Osada *et al*.[39] and Chen *et al*.[41].

In Fig. 7 is noted changes in the characteristics peaks. As the dopant content increases, the peaks at 627 $cm^{-1}$ shifted of about 20 $cm^{-1}$. We also observe a broadening of all the peaks when compared with La(0), as shown in Table 5.

In the studies of Refs.[39] and[41], the Raman modes at 460 $cm^{-1}$ and 630 $cm^{-1}$ can be associated with the vibrations of $O(3)_{Bi}A_{1g}$ and $O(2)_{Sr}A_{1g}$ along the *c* axis for the Bi-2212 phase. The $A_{1g}$ modes are symmetric for vibrations of Bi, Sr, Ca, Cu, $O_{Bi}$, $O_{Sr}$ and $O_{Cu}$ in the *c* axis, as reported by Carvalho and Guedes[40]. It is noted that the samples analyzed in this work exhibit similar behaviors to that one studied by Chen et al.[41], i.e., there are shifts of the Raman mode to lower values and the broadening of the peaks are due to the increasing of dopant contents. The study of Chen et al.[41] suggests that such results are consequences of the substitution of aliovalent ions in the sites of the Bi-2212 phase. Then, changes in the charge of around $O(2)_{Sr}$, alter the Sr-O(2) bond[39,41] and consequently the excess of oxygen caused by this substitution. This result is in agreement with the XRDs and electric measurements. The peak at around 540 $cm^{-1}$ for La(0) is not present in all samples. According to Carvalho and Guedes[40], the Raman mode in 530 $cm^{-1}$ is related to the $Ca_2PbO_4$, which is used to detect traces of this phase even for very small amounts of Pb. These peaks are more evident in La(0) and La(0.5).

## 4. Conclusion

We showed that the crystal structure and the electric, morphologic and magnetic properties of the BSCCO system are very sensitive to the doping process with La. The changes cause the degradation of the superconducting properties, which can be observed in the doped samples with contents above x = 0.5. This behavior is a consequence of the change in the charge carrier contents of the charge reservoirs and also the change in the conduction plans of this system. Those factors were caused by the substitution of Sr divalent ions ($Sr^{2+}$) by La trivalent ions ($La^{3+}$).

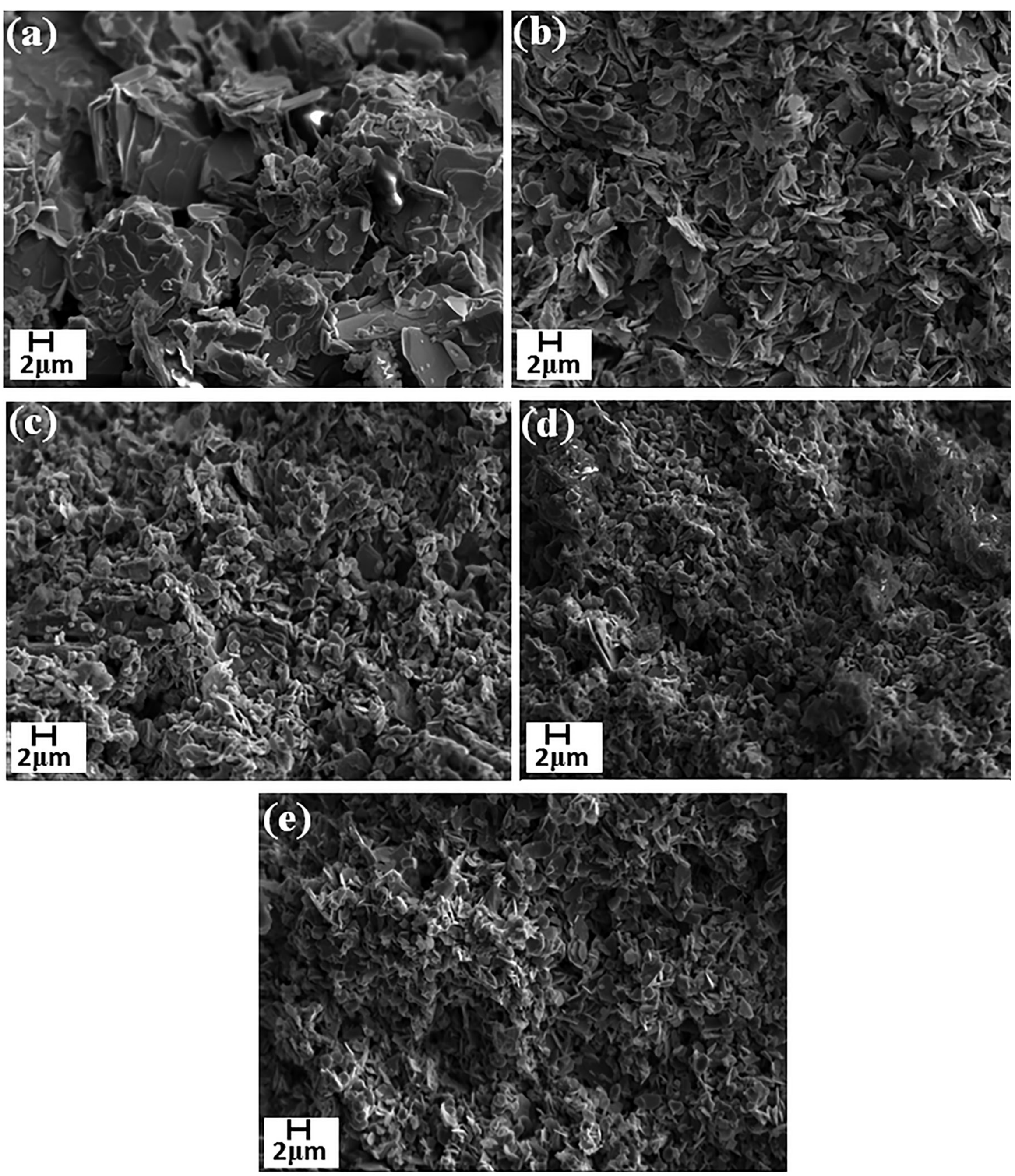

**Figure 5.** SEM micrographs of (a) La(0), (b) La(0.5),(c) La(1.0), (d)La(1.5) and (e)La(2.0) samples. The presence of plate-like grains is characteristic of the superconducting BSCCO system. By increasing the dopant content the sizes of the grain exhibit a more homogeneous distribution.

**Table 4.** Composition of the different samples analyzed by EDX.

| Sample | Average Composition |
|---|---|
| La(0) | $Bi_{1.52}Pb_{0.41}Sr_{1.90}Ca_{1.86}Cu_{3.30}O_y$ |
| La(0.5) | $Bi_{1.42}Pb_{0.42}Sr_{1.31}La_{0.57}Ca_{1.83}Cu_{3.45}O_y$ |
| La(1.0) | $Bi_{1.60}Pb_{0.31}Sr_{1.04}La_{1.11}Ca_{2.00}Cu_{2.94}O_y$ |
| La(1.5) | $Bi_{1.44}Pb_{0.33}Sr_{0.50}La_{1.72}Ca_{2.00}Cu_{3.11}O_y$ |
| La(2.0) | $Bi_{1.63}Pb_{0.27}La_{2.22}Ca_{1.94}Cu_{2.94}O_y$ |

## 5. Acknowledgments

We thank CAPES and CNPq for the financial support.

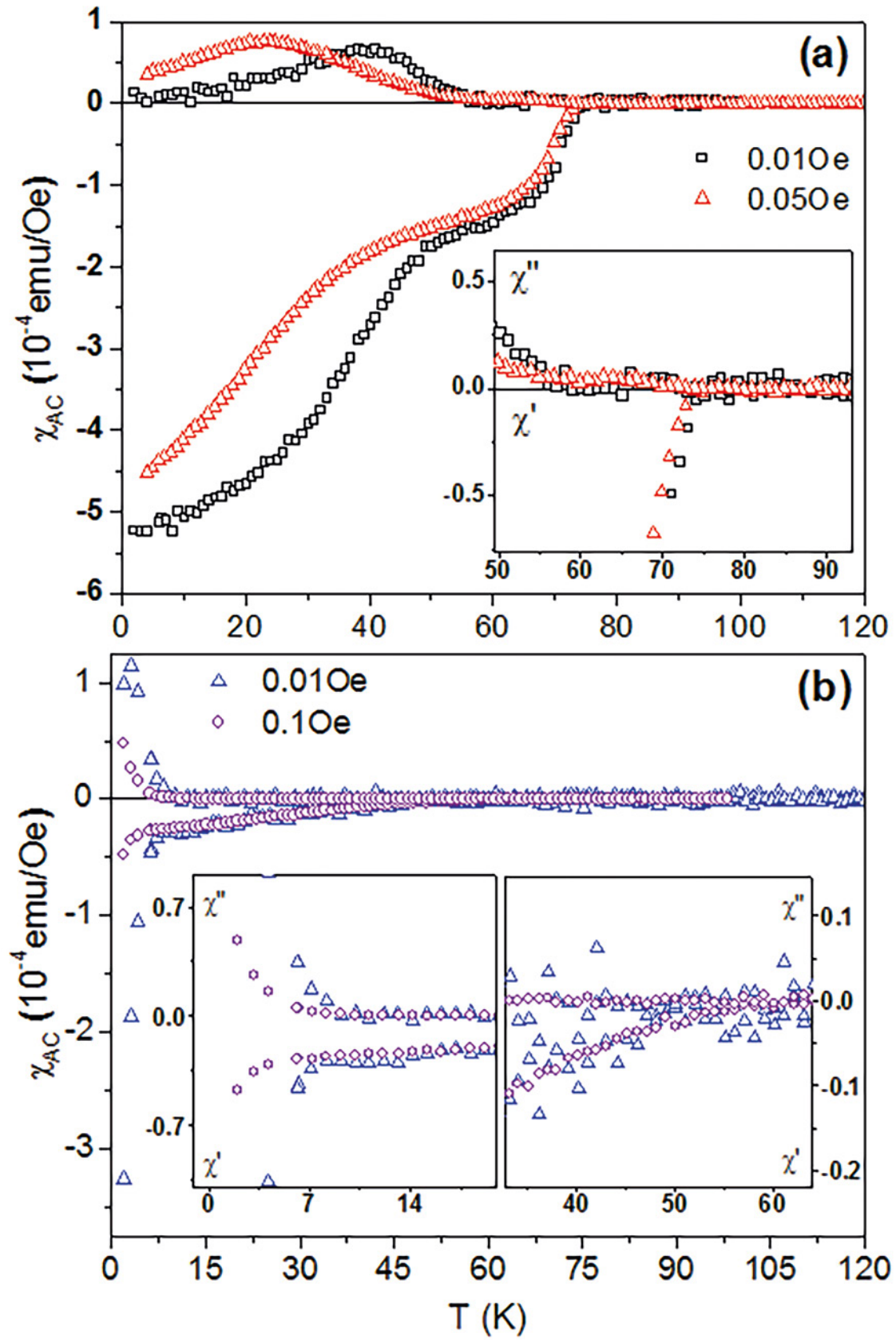

**Figure 6.** (a) AC susceptibility as a function of the temperature. (a) La(0) under an excitation field of 0.01 Oe and 0.05 Oe. (b) La(0.5) under an excitation field of 0.01 Oe and 0.1 Oe.

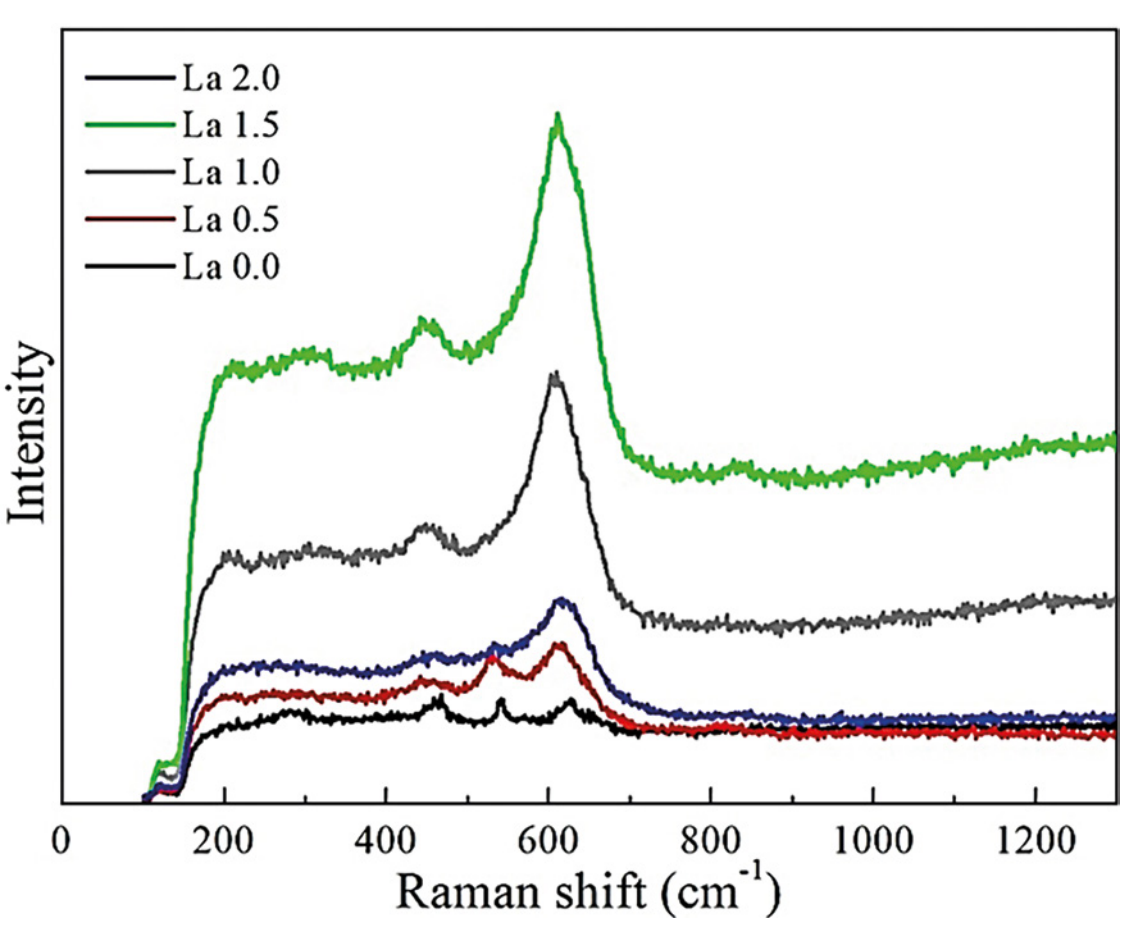

**Figure 7.** Raman spectra of the undoped and doped samples.

**Table 5.** Raman modes of the undoped and doped samples.

| La(0) ($cm^{-1}$) | La(0.5) ($cm^{-1}$) | La(1.0) ($cm^{-1}$) | La(1.5) ($cm^{-1}$) | La(2.0) ($cm^{-1}$) |
|---|---|---|---|---|
| 467 | 467 | 455 | 455 | 467 |
| 540 | 520 | | | |
| 627 | 613 | 609 | 609 | 613 |